\documentclass[longnamesfirst,apjl]{emulateapj}
\usepackage{apjfonts,natbib,epsfig}
\tighten

\newcommand{\beq}{\begin{equation}}
\newcommand{\eeq}{\end{equation}}
\newcommand{\beqa}{\begin{eqnarray}}
\newcommand{\eeqa}{\end{eqnarray}}
\def\msun{{\rm\,M_\odot}}
\def\zsun{{\rm\,Z_\odot}}
\def\etal{et al.\ }

\begin{document}

\title{
The Rise of Dwarfs and the Fall of Giants:  Galaxy Formation Feedback Signatures in the Halo Satellite Luminosity Function
}

\author{Asantha Cooray$^1$ and Renyue Cen$^2$}
\affil{$^1$Center for Cosmology, Department of Physics and Astronomy, University of California, Irvine, CA 92617\\
$^2$Princeton University Observatory, Princeton University, Princeton, NJ 08544.
}

\righthead{SATELLITE LUMINOSITY FUNCTION}

\lefthead{COORAY \& CEN}

\begin{abstract}
The observed luminosity function (LF) of satellite galaxies shows
several interesting features that require a better understanding of
gas-thermodynamic processes and feedback effects related to reionization and galaxy formation. In galaxy clusters, 
the abundance of dwarf galaxies is in good agreement with the expectation based on the subhalo mass function,
whereas in  galaxy groups, the relatively small abundance of dwarfs  
conflicts with theoretical expectations. In all halo systems, there is a dip in the abundance
of galaxies with luminosities in the range $\sim 2\times 10^8$ L$_{\sun}$ to $10^{10}$ L$_{\sun}$, corresponding
to subhalo mass scales between $\sim 5\times 10^{10}$ M$_{\sun}$ to few times $10^{11}$ M$_{\sun}$.
Photoionization from reionization has been used to explain statistics of the dwarf population, 
with larger systems forming prior to, and smaller systems forming subsequent to, reionization.
The observed dip in the LF is an imprint of small dwarf galaxies (<$\sim 2 \times 10^8$ L$_{\sun}$)
that formed prior to reionization. The galactic winds powered by supernovae in these dwarf galaxies propagate energy
and metals to large distances such that the intergalactic medium is uniformly enriched to a level of $10^{-3}\zsun$,
as observed in the low-redshift Ly$\alpha$ forest. 
The associated energy related to this metallicity  raises the intergalactic medium temperature and 
the Jeans mass to a range $10^{10}-10^{11}\msun$ at $z\sim 3.4-6.0$.
Because the epoch of nonlinearity for halos in this mass range
is at $z\ge 3.4-4.4$, their gas content, hence star formation, is greatly suppressed on average
and leads to a dip in the observed LF at $z=0$.
Larger halos ($M\ge 10^{11}\msun$), becoming nonlinear at $z\le 3.4-4.4$,
have masses topping the Jeans mass,
where subhalo mass function based LF is again in agreement with observations.


\end{abstract}

\section{Introduction}
\label{sec:introduction}

The complex physical processes  associated with reionization of the intergalactic medium (IGM) at a
redshift above 6 is
expected to leave characteristic scales and features in the star formation history (e.g., \citealt{Cen:03}),
in the supernovae distribution (e.g., \citealt{mesinger:05}), and, potentially, in the luminosity distributions of galaxies.
In the case of galaxy statistics, for example, 
the feedback related to supernovae heating in small mass halos and photoionization during reionization
\citep{Benson:02}. have been used to explain the flattened  faint-end slope of the galaxy luminosity function (LF). 

Since one averages galaxy statistics over large volumes and, thereby, averages 
over any inhomogeneities and differences in time scales
and dispersions coming from galaxy formation and evolution processes,
it is not possible to address detailed physics related to reionization with the galaxy LF alone.
On the other hand, the luminosity distribution of satellites in dark matter halos,
as a function of the halo mass, may be an ideal probe of reionization physics.  In this respect,
the lack of an abundant population of low luminosity galaxies in the local group, relative to
expectations from cold dark matter cosmological models, has been
explained in terms of photoionization \citep{Bul:00,Benson:03}. The subsequent {\it squelching} 
of galaxy formation in dark matter halos below a certain mass scale, corresponding to the temperature to which IGM
is heated, explains the environmental dependence of the faint-end slope of the cluster  LF
\citep{Tully:02}. 

In addition to effects related to photoionization, the cluster satellite LF should also show
signatures of feedback associated with starformation.
The satellite luminosity distribution can be described through the
halo occupation number \citep{Cooray:02}, minus the central galaxy, conditioned 
in terms of satellite luminosity. Here, we construct an empirical model for the 
conditional luminosity function (CLF; \citealt{Yang:03,Cooray:05,Cooray:05c})  of satellites in dark matter halos,
based on the subhalo mass function and 
compare to observed measurements; at the bright-end,
we make use of CLFs measured by \citet{Yang:05} using the 2dFGRS \citep{Cole:01} galaxy group catalog, and
extend this comparison to the faint-end using cluster LFs measured by \citet{Trentham:02} 
and \citet{Hilker:03}. We argue that cluster LFs show two scales, one associated with photoionization
at halo mass scales around $5 \times 10^{10}$ M$_{\sun}$, 
resulting in the disappearance of dwarfs satellites in galaxy groups relative to clusters, and another scale related to
an overall suppression of galaxy formation in subhalos below $10^{11}$ M$_{\sun}$, independent of the total system mass.

This {\it Letter} is organized as follows:
In \S~\ref{sec:LF}, we describe the construction of the satellite CLF of dark matter halos.  In \S~\ref{sec:discussion},
we compare our LF with the observed LFs of satellites in groups and clusters and discuss interesting physics that could
explain the observed features. We make use of the current 
concordance cosmological model \citep{Spergel:03}.

\section{Satellite Luminosity Function}
\label{sec:LF}

\begin{figure*}[!t]
\centerline{\psfig{file=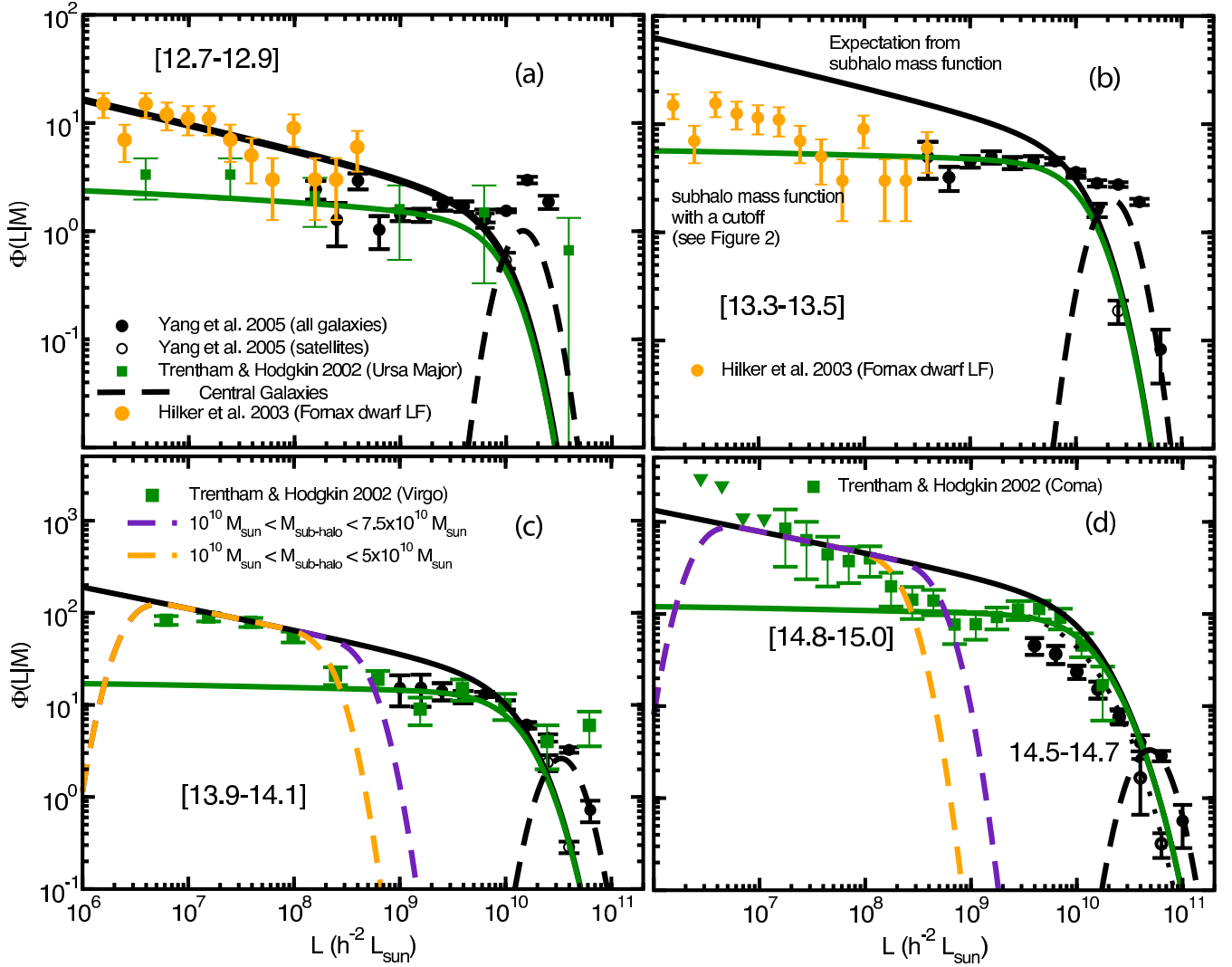,width=7.0in,angle=0}}
\caption{The CLF of dark matter halos as a function of the halo mass. 
From (a) to (d) we show four mass ranges from group to cluster scales,
 with  mass ranges (in logarithmic values) labeled within square brackets  in each of the figures.
The circles at the bright-end correspond to the measurements in the $b_J$-band by \citet{Yang:05}
using the 2dFGRS \citep{Cole:01} 
group catalog with a division to satellites (open circles) and satellites and central galaxies (solid circles). 
The long-dashed line is the central galaxy CLF following \citet{Cooray:05c}. The top solid line shows the expected CLF of satellites based on the subhalos mass function. The lower solid line is the satellite CLF when the subhalo mass function has a turnover at masses below 10$^{11}$ M$_{\sun}$. The faint-end data comes from
measurements in the B-band 
by \citet{Trentham:02} for Ursa Major, in (a) with a total mass around $\sim 8 \times 10^{12}$ M$_{\sun}$,
Virgo, in (b) with a total mass of $\sim 10^{14}$ M$_{\sun}$, and Coma, in (c), where we take the total mass to be  between 
 $(7$ and $10) \times 10^{14}$ M$_{\sun}$. In (a) and (b), we also show the faint-end LF of Fornax (from \citealt{Hilker:03}). In published literature, the total mass of Fornax varies between $(4.3$ to $8.1) \times 10^{12}$ M$_{\sun}$, based on 
X-ray data \citep{Jones:97}, to $\sim 7 \times 10^{13}$ (e.g., \citealt{Bekki:03}). If the low halo mass is correct for the
Fornax group, then Fornax formed earlier than Ursa Major  and before reionization was complete.
If the high mass is correct, given the presence of some dwarf galaxies, but below the level expected, 
Fornax probably was forming during reionization. The data can also be used to argue that 
the reionization process had inhomogeneities at size scales below that collapsed to form Fornax.
Note the difference in the y-axis scale between the two top panels and the two bottom panels.}
\label{fig:satLF}
\end{figure*}

The CLF, denoted by $\Phi(L|M)$, is the average number of galaxies with luminosities between $L$ and $L+dL$ that reside in halos of mass $M$ \citep{Yang:03,Cooray:05c}. Following \citet{Cooray:05},
 we separate the CLF into terms associated with central and satellite galaxies, $\Phi(L|M)=\Phi_{\rm c}(L|M)+\Phi_{\rm s}(L|M)$.
In previous studies \citep{Cooray:05c}, central galaxy CLF was described with a log-normal distribution in luminosity with 
a mean $L_{\rm c}(M)$ and dispersion $\Sigma_c$, while the satellite CLF is
assumed to be a power law. Here, we focus on the satellite CLF and, instead of an a priori assumption on a power-law
CLF, we model it  using the subhalo mass function.

In this approach, each satellite in a subhalo mass $M_s$ has a log-normal luminosity distribution of
\begin{eqnarray}
\phi_{\rm s}(L|M_s)  &=& \frac{ \phi(M_s)}{\sqrt{2 \pi} \ln(10)\Sigma L} \exp \left\{-\frac{\log_{10} [L /L_{\rm s}(M_s)]^2}{2 \Sigma^2}\right\}  \, ,
\end{eqnarray}
where the normalization $\phi(M_s)$ is fixed such that $\int\Phi(L|M_s)LdL=L_s(M_s)$.
Given the luminosity distribution of each satellite, the CLF of satellites, in a parent halo mass $M$,  is
\begin{equation}
\label{eq:lf}
\Phi_{\rm s}(L|M) = \int_0^\infty \phi_{\rm s}(L|M_s) \frac{dn_s(M_s|M)}{dM_s} dM_s \, ,
\end{equation}
where $dn_s(M_s|M)/dM_s$ is the subhalo mass function of dark matter halos given the parent halo mass $M$.
Here, we use the analytical form 
\begin{equation}
\frac{dn_s(M_s|M)}{dM_s} = \frac{\gamma}{\beta^2 M \Gamma(2-\alpha)} \left(\frac{M_s}{\beta M}\right)^{-\alpha} \exp \left( - \frac{M_s}{\beta M}\right) \, ,
\label{eqn:dns}
\end{equation}
where $\alpha=1.91$, $\beta=0.39$, and $\gamma=0.18$
\citep{Vale:04}. Our conclusions do not change significantly
if we use the description of  \citet{vanden:05} where $\alpha$  and $\gamma$ are functions of mass,
$M$, with $\alpha$ varying roughly over 10\% as $M$ is varied from group to cluster mass scales.
Since CLF measurements are averaged over a sample of dark matter halos in a narrow range in mass
\citep{Yang:05}, we also calculate the mass-averaged satellite CLF by averaging over the
dark matter halo mass function, $dn/dM$ (Sheth \& Tormen 1999), over the mass range of interest.

In our model, an important ingredient is the  $L_s(M_s)$ relation which describes the luminosity of a subhalo given the subhalo mass.
Here, we follow the procedure related to modeling the field galaxy LF \citep{Cooray:05}, and
employ a fitting function to describe the relation 
between the luminosity of a halo and the dark matter mass
of that halo. In \citet{Vale:04}, this relation was  obtained through a model description of the 2dFGRS LF \citep{Norberg:01} 
using the global subhalo mass function, $n_{\rm sh}(M_s) = \int dn/dM\; dn_s(M_s|M)/dM_s\; dM$.
The relation is 
\begin{equation}
\label{eq:fitting}
L(M) = L_0 \frac{(M/M_1)^{a}}{[b+(M/M_1)^{cd}]^{1/d}}\, .
\end{equation}
The relevant parameters for the $b_J$-band, as appropriate for 2dFGRS data, are $L_0=5.7\times10^{9} L_{\sun}$, $M_1=10^{11} M_{\sun}$,
$a=4.0$, $b=0.57$, $c=3.72$, and $d=0.23$ \citep{Vale:04}. 
Though this relation was used in
 \citet{Cooray:05c} to describe the field galaxy LF, whose statistics are dominated by central galaxies, the same relation should also
remain valid for subhalos as well. 
The remaining parameter in our model for the satellite LF  is $\Sigma$ and we set this to be 0.17 based on the value 
needed to explain the exponential drop-off in the field galaxy LF \citep{Cooray:05c}.

Note that at low subhalo masses, $dn_s/dM_s \propto M_s^{-\alpha}$ where $\alpha=1.91$,
independent of the halo mass, or varies from 2.0 to 1.9 when parent halo mass varies from $M \sim 10^{11}\msun$ to $10^{15}\msun$
\citep{vanden:05}. The satellite CLF is $\Phi_{\rm s}(L|M) \sim \int\phi_{\rm s}(L|M_s) (dn_s/dM_s)dM_s$.
If $L_{\rm s} \propto M_s^\eta$, we can write $\Phi_{\rm s}(L|M) \sim \int L^{\prime -1-\alpha/\eta +1/\eta} \delta(L'-L) dL'$,
where we have ignored the scatter in the $L_{\rm s}$--$M_s$ relation by setting $\Sigma\rightarrow0$.
The faint-end of the satellite LF then scales as $\Phi_{\rm s}(L|M) \propto L^{-1-\alpha/\eta+1/\eta}$.

\begin{figure}[!t]
\centerline{\psfig{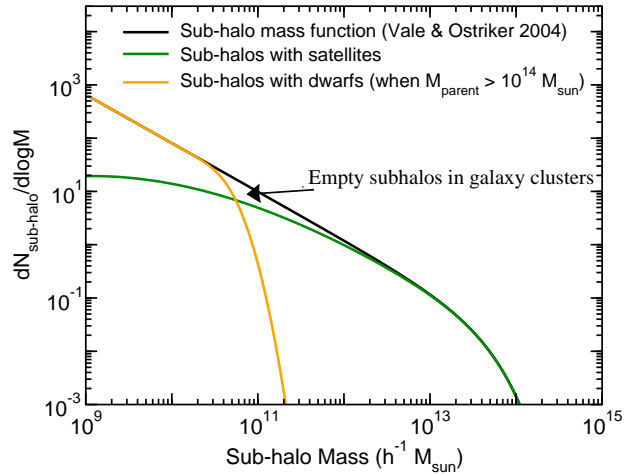}}
\caption{The subhalo mass function assuming a parent halo mass of $10^{14}$ M$_{\sun}$.
The solid line show the expectation based on CDM calculations (equation~\ref{eqn:dns}). 
The  middle line shows the flattening needed to
suppress the presence of galaxies fainter than $3 \times 10^9$ L$_{\sun}$ or the corresponding mass scale of
roughly $10^{11}$ M$_{\sun}$. The flattening is achieved through a smooth function discussed in Section~3.
The third line labeled ``sub-halos with dwarfs'' is the subhalo mass function necessary to explain the luminosity distribution at the faint-end of $L < 3 \times 10^8$ L$_{\sun}$ when the parent halo mass is above $10^{14}$ M$_{\sun}$.}
\label{fig:shmass}
\end{figure}

\section{Results and Discussion}
\label{sec:discussion}

In Figure~1, we show $b_J$-band CLFs based on the 2dFGRS group catalog \citep{Yang:05} and an extension to the faint-end based on nearby cluster LFs in the B-band. The satellite CLFs show several 
interesting trends. 
As one moves to a higher mass for the central halo, from (a) to (d) in Figure~1, one finds the faint-end of
the satellite CLF to be filled with dwarf galaxies. Since the 2dFGRS is incomplete below absolute magnitudes of
$M_{b_J}$ of -17, we make use of B-band LF measurements by \cite{Trentham:02}, whose
faint-end statistics are generally dominated by the dwarf galaxy population.
 At the high mass end of halos, corresponding to clusters like Virgo, with an assumed halo mass of
$10^{14}$ M$_{\sun}$, and Coma, with mass between $\sim (7$ and $10) \times 10^{14}$ M$_{\sun}$,
satellites begin to trace the expected slope predicted by the subhalo mass function.
On the other hand, in low mass groups, such as  Ursa
Major, with an assumed mass of $(5$ to $8) \times 10^{12}$ M$_{\sun}$, the faint-end dwarf population does not
trace the subhalo mass function. The statistics of the faint-end dwarf galaxy population
have been discussed in \citet{Tully:02} in the context of photoionization effect resulting from reionization. These
authors argue that massive clusters such as Virgo started to form prior to reionization, while low mass groups such as Ursa
Major, where the abundance of dwarf galaxies is relatively smaller, formed subsequent to reionization.  
As shown in Figure~1(c) and (d), the faint-end dwarf LF is associated with subhalo masses below $5 \times 10^{10}$ M$_{\sun}$.
The photoionized, and heated, gas is not expected to
 cool in halos of below this mass scale after reionization. Such an argument is consistent with Figure~1. 

Based on satellite LFs alone, we find that reionization may be more 
complex than simply assuming that the universe reionized completely at a single redshift.
For example, in Figure~1(a) and (b), we also show the LF of dwarf galaxies in Fornax. We plot these data
in both panels due to a large uncertainty, or variation, in the quoted total mass of Fornax in the literature.
Regardless of the exact mass of Fornax, the presence of more dwarfs than Ursa Major
suggests that Fornax formed prior to the latter system
and probably during reionization such that gas managed to cool in
a fraction of small dark matter halos while the rest was affected. The large scatter in dwarf population
of similar mass groups 
may be evidence for  inhomogeneous reionization or non-uniform feedback processes.

What is new and intriguing is this: when compared to the expectation based on the sub-halo mass function 
and the luminosity-mass relation of \citet{Vale:04},
one sees a relative decrease in the number of satellite galaxies at luminosities below $\sim 3 \times 10^9$ L$_{\sun}$,
corresponding to mass scales below few times $10^{11}$ M$_{\sun}$, in all halo systems.
Since at these mass scales, $3<\eta<4$, we expect $\Phi_{\rm s}(L|M)$ to scale as
$L^{-1.2}$ to $L^{-1.3}$; we do not expect the faint-end slope to be flatter than  $L^{-1.2}$, unless
the subhalo mass function slope is changed; for $\Phi_{\rm s}(L|M)$ to be luminosity independent, $\alpha \rightarrow 1$.
Thus, to suppress galaxies at luminosities below $3 \times 10^9$ L$_{\sun}$, we include an {\it efficiency} 
function to the subhalo mass function to characterize the subhalo mass distribution where  satellite galaxies present.
Here, we take an analytical description of the form $f(M_s) = 0.5( 1+ {\rm erf}\,[(\log M_s - \log M_c)]/\sigma)$,
 such that $f(M_s) \rightarrow 0$ when $M_s \gg M_c$ and $f(M_s) \rightarrow 0$ when $M_s \ll M_c$.
To explain the flattening of the satellite CLF, we set $M_c = 10^{11}$ M$_{\sun}$ and $\sigma=1.5$; with such a
broad dispersion, $f(M_s)\; dn(M_s|M)/dM_s$ flattens (see, Figure ~2) instead of becoming zero even when $M \sim 10^9$ M$_{\sun}$;
the flattening of the satellite CLF does not imply that all subhalos below some critical mass scale is affected, but one sees
a broad distribution of subhalo masses where, statistically, satellite 
galaxies are not present.  Instead of modifying the subhalo mass function, we can also vary the $L_s(M_s)$ relation
and set a steep slope for $\eta$.  While in this case all subhalos contain galaxies,  the average luminosity would be lower
for subhalos in groups relative to same mass subhalos in clusters. In such a scenario, it is also hard to understand
the sudden appearance of dwarfs in clusters, given the dip in the LF of giants regardless of system mass.

We now offer a physical explanation for this unique feature.
Supernovae-powered winds from abundant dwarf galaxies 
at $L<2\times 10^8$ M$_{\sun}$ are expected to be 
strong (Dekel \& Silk 1986; Mori et al. 2002).
We adopt the view that these winds transport
metals and energy into the IGM.
Following Cen \& Bryan (2001), the temperature of the IGM
is
\begin{eqnarray}
&&T_{SN}(z)= 1.3 \times 10^4 \; {\rm K} \;
	\left( \frac{E_{SN}}{1.2\times 10^{51}{\rm erg}} \right) \nonumber \\
&\times&	\left( \frac{M_C}{0.2\msun} \right)
	\left( \frac{\eta}{0.3} \right) 
	\left( \frac{[C/H]}{1\times 10^{-3}} \right)
	\left( \frac{4}{1+z} \right)^2 
\end{eqnarray}
where $M_C$ is the mass of carbon ejected by one supernova; 
$E_{SN}$ is the total
energy output of one supernova; $\eta$ is the fraction of that energy
that is eventually deposited in the IGM in the form of thermal energy
(Mori \etal 2002), and
$[C/H]$ is the ratio of carbon number density to hydrogen number density of the
gas in solar units.  
We have assumed that energy is deposited at some high redshift, perhaps $z\sim 6-9$,
and the temperature of the IGM subsequently decays adiabatically.
We adopt all the fiducial values for  $E_{SN}$, $M_C$, $\eta$ and $[C/H]$,
which gives us an added temperature to the IGM of $1.3\times 10^4$K
at $z \sim 3$; 
this explains the {\it Doppler width issue} of the Ly$\alpha$ forest (Cen \& Bryan 2001).

\begin{figure}[!t]
\centerline{\psfig{file=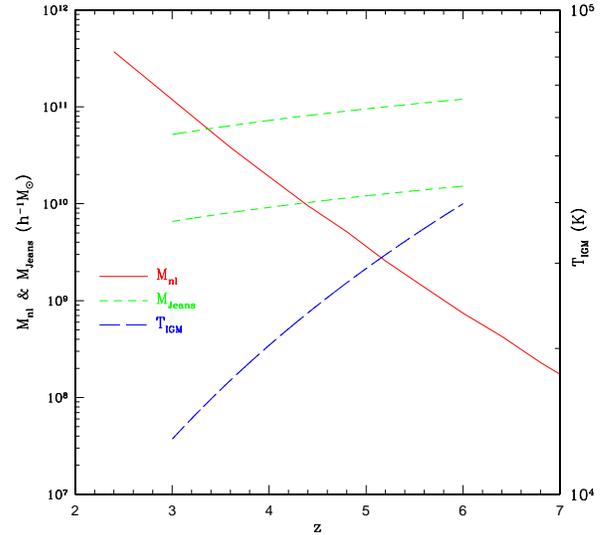,width=3.2in,angle=0}}
\caption{
The solid line shows the nonlinear mass scale
in the standard LCDM model as a function of redshift.
The two short-dashed curves show Jeans mass assuming
dark matter is uniformly distributed (upper curve) and
for gas at the virial density (i.e., 200 times the mean density) 
assuming gas and dark matter cluster in the same way (lower curve).
The long-dashed curve shows the evolution of IGM temperature (values labeled on the right y-axis).
}
\label{fig:jeans}
\end{figure}

Figure~3 shows the evolution of the nonlinear mass
in the standard cold dark matter model (solid curve),
the Jeans mass for two extreme conditions (two short dashed curves),
and the temperature of the IGM (long dashed curve).
We see that, depending on detailed gasdynamics,
the majority of halos formed at $z\ge 3.4-4.4$ and with
masses $10^{10}-6\times 10^{10}h^{-1}\msun$,
are significantly deprived of gas, potentially explaining the dip in the satellite LFs seen in Figure~1.

Our results suggest that most subhalos of 10$^{10}$ M$_{\sun}$ to $10^{11}$ M$_{\sun}$  in massive clusters
should not have galaxies in them (so-called ``dark halos''). Lensing-based studies indicate
the agreement between dark matter subhalo mass function 
and  the mass function of subhalos associated with galaxies in clusters
when $M_s > 3 \times 10^{11}$ M$_{\sun}$ \citep{Natarajan:04}.
While the observed difference below this mass scale is considered to be
due to an observational limitation, it is useful to extend lensing studies to measure the subhalo mass function below
$10^{11}$ M$_{\sun}$. The dip in the LF could be detected as a dip in the mass function of subhalos with
galaxies (Figure~2). Furthermore, 
a large sample of satellite LFs, down to dwarf galaxy magnitudes and with better determined parent halo masses,
have the potential to address more  detailed physics of reionization and galaxy formation processes 
through a better characterization of features in the CLF.

\acknowledgements
AC thanks Frank van den Bosch for helpful correspondence and X. Yang for LF measurements.
This work was completed while AC was at the Aspen Center for Physics.
This work is supported in part by grants AST-0206299, AST-0407176
and NAG5-13381 to RC.

\end{document}